\documentclass{elsart}
\usepackage{natbib}
\usepackage{harvard}

\def\ZZ#1{$\scriptstyle #1$}
\newcommand{\rn}[1]{(\ref{#1})}
\newcommand{\be}{\begin{equation}}
\newcommand{\ee}{\end{equation}}
\newcommand{\bea}{\begin{eqnarray}}
\newcommand{\eea}{\end{eqnarray}}
\begin{document}
\begin{frontmatter}
\title{\bf NBODY2: A DIRECT N-BODY INTEGRATION CODE}
\author{ Sverre J. Aarseth}
\address{Institute of Astronomy, Madingley Road, Cambridge CB3 0HA,
 England}
\date{ \today}
\begin{abstract}
We give a full description of the code \ZZ {NBODY2} for direct
integration of the gravitational $N$-body problem.
The method of solution is based on the neighbour scheme of Ahmad \& Cohen
(1973) which speeds up the force calculation already for quite modest
particle numbers.
Derivations of all the relevant mathematical expressions are given,
together with a detailed discussion of the algorithms.
The code may be used to study a wide variety of self-consistent problems
based on a small softening of the interaction potential.
\end{abstract}

\thanks[sja] {E-mail: sverre@ast.cam.ac.uk}
\end{frontmatter}

\section{Introduction}
$N$-body simulations provide a useful tool for exploring a wide
range of astronomical problems.
By now, direct summation methods have been used to study such diverse
topics as planetary formation, few-body scattering, star cluster dynamics,
violent relaxation, tidal disruption of dwarf galaxies, interacting
galaxies, compact groups and clusters of galaxies, as well as cosmological
modelling.
The code \ZZ {NBODY2} described here is suitable for relatively small
$N$-body systems ($20 < N < 10^4$).
However, it does not contain any special treatments of close encounters
and should therefore only be used for problems where the effect of
hard binaries can be neglected (a variety of regularization methods are
discussed in Aarseth 1985).

\section{Integration Method}
In this Section, we provide all the relevant mathematical expressions 
and discuss the basic principles of $N$-body integration.

\subsection{Difference Formulation}
We write the equation of motion for a particle of index $i$ in the form
\be
   \ddot{ {\bf r}}_i \, = 
\, -G \sum_{j=1;\, j \not = \,i}^N {m_j \,({\bf r}_i - {\bf r}_j)
 \over {(r_{ij}^2 + \epsilon^2)^{3/2} } }. \label{newton}
\ee
Here $G$ is the gravitational constant and the summation is over all the
other particles of mass $m_j$ and coordinates ${\bf r}_j$.
The introduction of a softening parameter $\epsilon$ prevents a force
singularity as the mutual separation
$r_{ij} \to 0$, thereby making the numerical solutions well behaved.
For convenience, we adopt scaled units in which $G = 1$ and define the
left-hand side of \rn{newton} as the force {\it per unit mass}, ${\bf F}$,
omitting the subscript.
The present difference formulation is based on the notation of
Ahmad \& Cohen (1973, hereafter AC) and follows closely an earlier
treatment (Aarseth 1985).

   Given the values of ${\bf F}$ at four successive past epochs
$t_3,\, t_2,\, t_1,\, t_0$, with $t_0$ the most recent, we write a
fourth-order fitting polynomial at time $t$ as
\be
{\bf F}_t = \left( \left ( ({\bf D}^4 (t - t_3)  +
{\bf D}^3 \right ) (t - t_2) + {\bf D}^2 \right ) (t - t_1) +
{\bf D}^1) (t - t_0) + {\bf F}_0. \label{Ft}
\ee
Using compact notation, the first three divided differences are defined by
\be
{\bf D}^k[t_0,t_k] \, = \, { {\bf D}^{k-1}[t_0,t_{k-1}] \,-\, 
{\bf D}^{k-1}[t_1,t_k]
\over {t_0 - t_k} }, ~~~~~~~(k = 1,\, 2,\, 3)  \label{diff}
\ee
where ${\bf D}^0 \equiv {\bf F}$
and square brackets refer to the appropriate time intervals
(${\bf D}^2 [t_1,t_3]$ is evaluated at $t_1$).
The term ${\bf D}^4$ is defined similarly by ${\bf D}^3[t,t_2]$ and 
${\bf D}^3[t_0,t_3]$.
Conversion of the force polynomial into a Taylor series provides simple
expressions for integrating coordinates and velocities.
Equating terms in the successive time derivatives of \rn{Ft} with an
equivalent Taylor series and setting $t = t_0$ yields the force derivatives
\bea
{\bf F}^{(1)} \, &=& \, \left( ({\bf D}^4 t'_3 \, + \, {\bf D}^3) \,
t'_2 \, + \, {\bf D}^2 \right ) \, t'_1 \, + \, {\bf D}^1 \nonumber \\
{\bf F}^{(2)} \, &=& \, 2!\, \left ( {\bf D}^4(t'_1 t'_2 + t'_2 t'_3 + t'_1 t'_3) \,
+ \, {\bf D}^3(t'_1 + t'_2) \, + \, {\bf D}^2 \right ) \nonumber \\
{\bf F}^{(3)} \, &=& \, 3!\, \left ( {\bf D}^4(t'_1 \, + \, t'_2 \, + \, t'_3) \, + \,
{\bf D}^3 \right ) \nonumber \\
{\bf F}^{(4)} \, &=& \, 4!\, {\bf D}^4, \label{taylor}
\eea
where $t'_k = t_0 - t_k$.
Equation \rn{taylor} is mainly used to obtain the Taylor series
derivatives at $t = t_0$, when the fourth difference is not yet known.
Thus the contribution from ${\bf D}^4$ to each order is only added at the
end of an integration step.
This so-called `semi-iteration' gives increased accuracy at little extra cost
(on scalar machines) and no extra memory.

   We now describe the initialization procedure, assuming one force polynomial.
From the initial conditions $m_j,\, {\bf r}_j,\, {\bf v}_j$, the respective
Taylor series derivatives are formed by successive differentiations of
\rn{newton}.
Introducing the relative coordinates
${\bf R} = {\bf r}_i - {\bf r}_j$ and the relative velocities
${\bf V} = {\bf v}_i - {\bf v}_j$,
all pair-wise interaction terms in ${\bf F}$ and ${\bf F}^{(1)}$
are first obtained by
\bea
{\bf F}_{ij} \, &=& \, -{ m_j \, {\bf R} \over R^3} \cr
{\bf F}^{(1)}_{ij} \, &=& \, -{m_j \, {\bf V} \over R^3 } - 3 a \, {\bf F}_{ij},
\label{fij}
\eea
with $a = {\bf R} \cdot {\bf V} / R^2$.
The total contributions are obtained by summation over $N$.
Next, the mutual second- and third-order terms are formed from
\bea
{\bf F}_{ij}^{(2)} \, &=& \, -{{ m_j \, ({\bf F}_i - {\bf F}_j)} \over R^3}
\, - \, 6 a \, {\bf F}_{ij}^{(1)} \, - \, 3 b \, {\bf F}_{ij} \nonumber \\
{\bf F}_{ij}^{(3)} \, &=& \, -{{ m_j \, ({\bf F}_i^{(1)} - {\bf F}^{(1)}_j)}
 \over R^3} \, - \, 9 a \, {\bf F}_{ij}^{(2)} \, -
 \, 9 b \, {\bf F}_{ij}^{(1)} - \, 3 c \, {\bf F}_{ij},  \label{f2ij}
\eea
with
\bea
b \, &=& \, ({V \over R})^2 \, + \, {{ {\bf R} \cdot ({\bf F}_i - {\bf F}_j)}
\over R^2} \, + \, a^2 \nonumber \\
c \, &=& \, {{ 3 {\bf V} \cdot ({\bf F}_i - {\bf F}_j)} \over R^2}  \, + \,
{{{\bf R} \cdot ({\bf F}^{(1)}_i - {\bf F}^{(1)}_j}) \over R^2 }
\, + a \, (3 b - 4 a^2).
\eea
A second double summation then gives the corresponding values of ${\bf F}^{(2)}$
and ${\bf F}^{(3)}$ for all particles.
This pair-wise boot-strapping procedure provides a convenient starting
algorithm, since the extra cost is usually small.

   Appropriate initial time-steps $\Delta t_i$ are now determined, using the
general criterion discussed in the next subsection.
Setting $t_0 = 0$, the backward times are initialized by 
$t_k = -k \, \Delta t_i \, \,(k = 1,\, 2,\, 3)$.
Inversion of \rn{taylor} to third order yields starting 
values for the divided differences,
\bea
{\bf D}^1 \, &=&
\, ( {1 \over 6} \, {\bf F}^{(3)} \,t'_1  \,
- \, {1 \over 2} \, {\bf F}^{(2)} ) \,
t'_1 \, + \, {\bf F}^{(1)} \nonumber \\
{\bf D}^2 \, &=& \,- {1 \over 6}  \,{\bf F}^{(3)}  \,(t'_1 + t'_2) \, + \,
{1 \over 2} \, {\bf F}^{(2)} \nonumber \\
{\bf D}^3 \, &=& \, {1 \over 6} \, {\bf F}^{(3)}. \label{Ddiff}
\eea

\subsection{Individual Time-Steps}
Stellar systems are characterized by a range in density which gives rise
to different time-scales for significant changes of the orbital parameters.
In order to exploit this feature, and economize on the expensive force
calculation, each particle is assigned its own time-step which is related
to the orbital time-scale.
Thus the aim is to ensure the convergence of the force polynomial \rn{Ft}
with the minimum number of force evaluations.
Since all interactions must be added consistently in a direct integration
method, it is necessary to include a temporary coordinate prediction of the
other particles.
However, the additional cost of low-order predictions still leads to a
significant saving since this permits arbitrarily large time-step ratios.

   Following the polynomial initialization discussed above, the integration
cycle itself begins by determining the next particle ($i$) to be advanced.
Thus in general, $i = \min_j \,(t_j + \Delta t_j)$, where $t_j$ is the time
of the last force evaluation.
It is convenient to define the present epoch (or `global' time) by
$t = t_i + \Delta t_i$, rather than relating it to the previous value.

   The individual time-step scheme (Aarseth 1963) uses two types of co\-ordinates
for each particle. We define primary and secondary coordinates, ${\bf r}_0$ and
${\bf r}_t$, evaluated at $t_0$ and $t$, respectively, where the latter are
derived from the former by the predictor.
In the present treatment where high precision is not normally required, we
predict coordinates to order $\bf {F}^{(1)}$ by
\be
{\bf r}_t \, = \, ( ( {\bf \widetilde F}^{(1)} \, \delta t'_j \,
+ \,  {\bf \widetilde F} ) \,
\delta t'_j \, + \, {\bf v}) \, \delta t'_j \, + \, {\bf r}_0,  \label{pred}
\ee
where ${\bf \widetilde F}^{(1)} = {1 \over 6} {\bf F}^{(1)}, \,
{\bf \widetilde F} = {1 \over 2} {\bf F}$ and $\delta t_j' = t - t_j$
(with $\delta t_j' < \Delta t_j$).
The coordinates and velocities of particle $i$ are then improved
to order ${\bf F}^{(3)}$ by standard Taylor series integration (cf.~\rn{taylor}),
whereupon the current force may be obtained by direct summation.
At this stage the four times $t_k$ are updated to be consistent with the
definition that $t_0$ denotes the time of the most recent force evaluation.
New differences are now formed (cf.~\rn{diff}), including ${\bf D}^4$.
Together with the new ${\bf F}^{(4)}$, these correction terms are 
combined to improve the current coordinates and velocities to highest order,
whereupon the primary coordinates are initialized by ${\bf r}_0 = {\bf r}_t$.

   New time-steps are assigned initially for all particles and at the end
of each integration cycle for particle $i$.
We adopt the composite criterion
\be
\Delta t_i \, = \, \left ( {\eta \,(\vert {\bf F}\vert \,
\vert {\bf F}^{(2)} \vert \, +
\, \vert {\bf F}^{(1)} \vert ^2) } \over
{ (\vert {\bf F}^{(1)} \vert \, \vert {\bf F}^{(3)} \vert \, +
\, \vert {\bf F}^{(2)} \vert ^2) } \right ) ^{1/2}, \label{step}
\ee
where $\eta$ is a dimensionless accuracy parameter.
For this purpose, only the last two terms of the first and second force
derivatives in \rn{taylor} are included.
This expression ensures that all the force derivatives play a role and
is also well defined for special cases (i.e. starting from rest
or $\vert {\bf F} \vert \simeq 0$).
Although successive time-steps normally change smoothly, it is prudent to
restrict the growth by a stability factor (presently 1.2).

In summary, the scheme requires the following 30 variables for each particle:
$m,\, {\bf r}_0,\, {\bf r}_t,\, {\bf v}_0,\, {\bf F},\, {\bf F}^{(1)},\,
{\bf D}^1,\, {\bf D}^2,\, {\bf D}^3,\, \Delta t,\, t_0,\, t_1,\, t_2,\, t_3$.
It is also useful to employ a secondary velocity, ${\bf v}_t$,
for dual purposes.

\subsection{Neighbour Scheme}
Having introduced the basic tools for direct integration, we now turn to
the AC neighbour scheme.
Here the main idea is to reduce the effort of evaluating the force contribution
from distant particles by employing two polynomials based on separate
time-scales.
Splitting the total force on a given particle into an irregular and a regular
component,
\be
{\bf F} = {\bf F}_{\rm irr} + {\bf F}_{\rm reg},  \label{ftot}
\ee
we can replace the full $N$ summation in \rn{newton} by a sum over the $n$
nearest particles together with a prediction of the distant contribution.
This procedure can lead to a significant gain in efficiency, provided the 
respective time-scales are well separated and $n << N$.

   To implement the AC scheme, we form a list for each particle containing
all members inside a sphere of radius $R_s$.
In addition, we include any particles within a
surrounding shell of radius $2^{1/3} \, R_s$ satisfying
${\bf R} \cdot {\bf V} <  0.1 \, R^2_s / \Delta T_i $,
where $\Delta T_i$ denotes the regular time-step.
This ensures
that fast approaching particles are selected from the buffer zone.

   The size of the neighbour sphere is modified at the end of each regular
time-step when a total force summation is carried out.
We have adopted a neighbour criterion mainly based on distance for
convenience.
However, some modification would be desirable for very large mass ratios
(say $> 100$).
A selection criterion based on the local number density contrast has proved
itself for a \hbox {variety} of problems, but may need modification for
interacting subsystems or massive particles.
To sufficient approximation, the local number density contrast is given by
\be
C = {2 \, n_1 \over N } \left ( {R_h \over R_s } \right )^3,  \label{contr}
\ee
where $n_1$ is the current membership and $R_h$ is the half-mass radius.
In order to limit the range, we adopt a predicted membership
\be
n_p =  n_{\rm max} \, (0.04 \, C)^{1/2},  \label{np}
\ee
subject to $n_p$ being within [$0.2 \, n_{\rm max}, \, 0.9 \, n_{\rm max}$],
with $n_{\rm max}$ denoting the maximum permitted value.
The new neighbour sphere radius is then adjusted using the corresponding
volume ratio, which gives
\be
R^{\rm new}_s = R^{\rm old}_s \, \left ( {n_p \over n_1 } \right )^{1/3}.
\label{rsnew}
\ee

   An alternative and simpler strategy is to stabilize all neighbour memberships
on the same constant value $n_p = n_0$ (AC, Makino \& Hut 1988).
A number of refinements are also included as follows:

\begin{itemize}
\item {In order to avoid a resonance oscillation in $R_s$,
$\, (n_p / n_1)^{1/6}$ is used in \rn{rsnew} if the predicted
membership lies between the old and new value}
\item {$R_s$ is modified by a radial velocity factor
outside the core}
\item {The volume factor is only allowed to change by 25 \%, subject
to a time-step dependent cut-off if $\Delta T_i < 0.01 \, T_{\rm cr}$
($T_{\rm cr}$ is the crossing time)}
\item {If $n_1 \leq 3$ and the neighbours are moving outwards, the
standard sphere $R_s$ is increased by 10~\%.}
\end{itemize}

   The gain or loss of particles is recorded when comparing the old and new
neighbour list, following the re-calculation of the regular force.
Regular force differences are first evaluated, assuming there has been no
change of neighbours.
This gives rise to the provisional new regular force difference
\be
{\bf D}^1 \,= \, { {\bf F}^{\rm new}_{\rm reg} \, - \,
({\bf F}^{\rm old}_{\rm irr} - {\bf F}^{\rm new}_{\rm irr}) \,
- \, {\bf F}^{\rm old}_{\rm reg}
\over { t - T_0}}, \label{Dreg}
\ee
where ${\bf F}^{\rm old}_{\rm reg}$ denotes the old regular force,
evaluated at time $T_0$, and the net change in the irregular force is
contained in the middle brackets.
In the subsequent discussions, regular times and time-steps are
distinguished using upper case characters.
All current force components are obtained using the {\it predicted}
coordinates (which must be saved), rather than the corrected values based on
the irregular ${\bf D}^4$ term since otherwise \rn{Dreg} would contain a
spurious force difference.
The higher differences are formed in the standard way, whereupon the
regular force corrector is applied (if desired).

   A complication arises because any change in neighbours requires
appropriate corrections of both force polynomials; i.e. using the
principle of successive differentiation of \rn{ftot}.
The respective Taylor series derivatives \rn{fij} and \rn{f2ij} are accumulated
to yield the net change.
Each force polynomial is modified by first adding or subtracting the
correction terms to the corresponding Taylor series derivatives \rn{taylor}
(without the ${\bf D}^4$ term), followed by a conversion to standard
differences using \rn{Ddiff}.

   Implementation of the AC scheme requires the following additional set of
regular variables: ${\bf F}_{\rm reg},\, {\bf D}^1,\, {\bf D}^2,\, {\bf D}^3,\,
\Delta T,\, T_0,\, T_1,\, T_2,\, T_3$, as well as the neighbour sphere radius
$R_s$ and neighbour list $L$ (size $n_{\rm max} + 1$).

\section{ Program Structure}
This Section describes the structure of the code and defines the main
input parameters and some of the options.

\subsection{Routines}
The program contains some 2307 lines of Fortran statements
and another 1203 lines of comments and spaces.
It is divided into 38 routines, many of which are optional and not
required for standard simulations.
All the routines are listed in Table~1, together with the
main calling sequence and an outline of the purpose (\# denotes an option).

\medskip
\begin{table}
\caption{Names of routines}
\renewcommand{\arraystretch}{1}
\begin{tabular}{lll}
\hline
 Routine &Called by &Description  \\
\hline
\ZZ {BLOCK} &        &Block data initialization \\
\ZZ {BODIES} &\ZZ {OUTPUT} &Output of binaries and particles (\# 6 \& 9) \\
\ZZ {CHECK}  &\ZZ {OUTPUT} &Error control and restart (\# 2 \& 11) \\
\ZZ {CMCORR} &\ZZ {OUTPUT} &Centre of mass corrections (\# 18) \\
\ZZ {COAL}  &\ZZ {MAIN}  &Inelastic two-body collision (\# 12) \\
\ZZ {CORE}   &\ZZ {OUTPUT} &Core radius and density centre (\# 8) \\
\ZZ {CPUTIM} &\ZZ {OUTPUT} \& \ZZ {INTGRT} &CPU time in minutes
 (machine dependent) \\
\ZZ {DATA}   &\ZZ {START}  &Generation of initial conditions \\
\ZZ {DEFINE} &\ZZ {INPUT}  &Definition of input parameters and counters \\
\ZZ {ENERGY} &\ZZ {OUTPUT} \& \ZZ {SCALE} &Total energy \\
\ZZ {ESCAPE} &\ZZ {OUTPUT} &Removal of escaping particles (\# 13) \\
\ZZ {FPCORR} &\ZZ {REGINT} &Derivative corrections of force polynomials \\
\ZZ {FPOLY1} &\ZZ {START} \& \ZZ {COAL} &Force and first derivative \\
\ZZ {FPOLY2} &\ZZ {START} \& \ZZ {COAL} &Second and third force derivatives \\
\ZZ {INPUT}  &\ZZ {START}  &Main parameter input \\
\ZZ {INSERT} &\ZZ {INTGRT} &Insert of particle index in sequential list \\
\ZZ {INTGRT} &\ZZ {MAIN}   &Decision-making and flow control \\
\ZZ {LAGR}   &\ZZ {OUTPUT} &Lagrangian radii and half-mass radius (\# 7) \\
\ZZ {MAIN}  &        &Master control flow \\
\ZZ {MODIFY}  &\ZZ {MAIN}  &Reading modified parameters at restart \\
\ZZ {MYDUMP} &\ZZ {MAIN} \& \ZZ {INTGRT} &\ZZ {COMMON} save or copy (\# 1 \& 2) \\
\ZZ {NBINT}  &\ZZ {INTGRT} &Irregular integration \\
\ZZ {NBLIST} &\ZZ {START} \& \ZZ {COAL}  &Initialization of neighbour list \\
\ZZ {OUTPUT} &\ZZ {MAIN}   &Main output and data save \\
\ZZ {RAN2}  &\ZZ {DATA}  &Portable random number generator \\
\ZZ {REGINT} &\ZZ {INTGRT} &Regular integration \\
\ZZ {REMOVE} &\ZZ {ESCAPE} \& \ZZ {COAL} &Removal of particle from
 \ZZ {COMMON} arrays \\
\ZZ {SCALE} &\ZZ {START} &Scaling to new units \\
\ZZ {SORT2}  &\ZZ {INTGRT} \& \ZZ {LAGR} &Sequential sorting of array \\
\ZZ {START}  &\ZZ {MAIN}  &Calling sequence for initial setup \\
\ZZ {STEPI}  &\ZZ {INTGRT} &Time-step for irregular force polynomial \cr
\ZZ {STEPS}   &\ZZ {START} \& \ZZ {COAL} &Initialization of time-steps and
 differences \\
\ZZ {SUBSYS}  &\ZZ {START}  &Initial subsystems for binary test (\# 17) \\
\ZZ {VERIFY}  &\ZZ {INPUT}  &Validation of main input parameters \\
\ZZ {XTRNL1} &\ZZ {REGINT} \& \ZZ {FPOLY1} &Force due to Plummer
 potential (\# 15) \\
\ZZ {XTRNL2} &\ZZ {REGINT} \& \ZZ {FPOLY1} &Force due to logarithmic
 potential (\# 15) \\
\ZZ {XVPRED} &\ZZ {OUTPUT} \& \ZZ {COAL}  &Prediction of coordinates and
 velocities \\
\ZZ {ZERO}  &\ZZ {START}  &Initialization of global scalars \\
\hline
\end{tabular}
\end{table}
\medskip

   The program consists of four separate parts: input, output,
initialization and integration.
The basic routines are: \ZZ {START, ~INPUT, ~DATA, ~SCALE} (input);
\ZZ {ENERGY}, \ZZ {CORE}, \ZZ {OUTPUT} (output);
\ZZ {NBLIST}, \ZZ {FPOLY1}, \ZZ {FPOLY2}, \ZZ {STEPS} (initialization);
and \ZZ {INTGRT}, \ZZ {NBINT}, \ZZ {REGINT}, \ZZ {STEPI} (integration).
The main calling sequence (\ZZ {START}, \ZZ {OUTPUT}, \ZZ {INTGRT})
originates in routine \ZZ {MAIN}, whereas routine \ZZ {START} acts as 
driver for both input and polynomial initialization and routine
\ZZ {INTGRT} controls the subsequent program flow.

To improve the layout
and distinguish between code and text, we have adopted upper case characters
for the former.
At the `microscopic' level, blank spaces are used to improve
readability of the Fortran statements, where blocks of code inside
\ZZ {DO}-loops and logical \ZZ {IF}'s are indented.

\subsection{COMMON Variables}
In principle, the program can be used for any particle number,
subject to available computer memory and CPU time.
The size of all particle arrays is controlled by the parameter \ZZ {NMAX},
together with \ZZ {LMAX} which is used for the neighbour lists.
These parameters must be specified in the auxiliary header file {\tt params.h}
at compile time since most routines communicate via the global \ZZ {COMMON}
blocks, rather than by general-purpose arguments.
All the \ZZ {COMMON} variables are defined in a \TeX\ file for
convenience.\footnote{The code can be downloaded from
http://www.ast.cam.ac.uk/$\sim$sverre.}

   The global labelled \ZZ {COMMON} blocks (header file {\tt common2.h})
are organized in four separate parts denoted
\ZZ {NBODY, ~NAMES, ~COUNTS, ~PARAMS}, which contain all the essential variables.
This enables a calculation to be halted temporarily by saving the \ZZ {COMMON}
blocks on disc or tape, whereupon a restart can be made directly from the
high-order scheme.

   The convention of up to six characters for the names of
routines or variables has been adopted, using mnemonic definitions.
Standard Fortran 77 has been adhered to as far as possible.
The main exceptions are: 

\begin{itemize}
\item {The \ZZ {INCLUDE} statement is used for \ZZ {COMMON} blocks}
\item {Input data are read in free format, which is less
prone to typing errors}
\item {Neighbour lists are declared as \ZZ {INTEGER}$\ast2$ to save memory}
\item {Continuation lines have `$\&$' in column 6 (non-standard character).}
\end{itemize}

\subsection{Input Parameters}
To be flexible and useful, a general-purpose program should allow a choice
of input data.
Since parameter space is large, the permissible range cannot be specified
uniquely in advance and every new simulation becomes a test case.
The complete list of input parameters is defined in Table~2,
where each input record is separated by a space.
The corresponding list of options (denoted \ZZ {KZ(J))} can be found in
routine \ZZ {DEFINE}.

\medskip
\begin{table}
\caption{Input parameters}
\renewcommand{\arraystretch}{1}
\begin{tabular}{lll}
\hline
 Variable &Definition \\
\hline
\ZZ {KSTART}  &Control index (1: new run; $>$1: restart; $>2$: new parameters) \\
\ZZ {TCOMP}   &Maximum computing time in minutes (saved in \ZZ {CPU})\\~\\
\ZZ {N}       &Total particle number ($\le$ \ZZ {NMAX}) \\
\ZZ {NFIX}    &Output frequency of data save or binaries (\# 3 \& 6) \\
\ZZ {NRAND}   &Random number sequence skip \\
\ZZ {NNBMAX}  &Maximum number of neighbours ($<$ \ZZ {LMAX}) \\
\ZZ {NRUN}    &Run identification index \\~\\
\ZZ {ETAI}    &Time-step parameter for irregular force polynomial \\
\ZZ {ETAR}    &Time-step parameter for regular force polynomial \\
\ZZ {RS0}     &Initial neighbour sphere radius \\
\ZZ {DELTAT}  &Output time interval in units of the crossing time \\
\ZZ {TCRIT}   &Termination time in units of the crossing time \\
\ZZ {QE}      &Energy tolerance (restart if \ZZ {DE/E} $> 5\ast$\ZZ {QE}
              and \ZZ \# 2 $>$ 1) \\
\ZZ {EPS}  &Softening parameter (square saved in \ZZ {EPS2}) \\~\\
\ZZ {KZ(J)}   &Non-zero options for alternative paths \\~\\
\ZZ {XTPAR1}  &Mass of external Plummer model (\ZZ \# 15 = 1; scaled units) \\
\ZZ {XTPAR2}  &Length scale for Plummer model \\~\\
\ZZ {ZMGAS}   &Mass scale for external logarithmic potential (\ZZ \# 15 = 2) \\
\ZZ {RGAS}    &Length scale for logarithmic potential \\~\\
\ZZ {ALPHAS}  &Power-law index for initial mass function (routine \ZZ {DATA}) \\
\ZZ {BODY1}   &Maximum particle mass before scaling \\
\ZZ {BODYN}   &Minimum particle mass before scaling \\~\\
\ZZ {Q}       &Virial ratio (routine \ZZ {SCALE}; \ZZ {Q} = 0.5 for equilibrium) \\
\ZZ {VXROT}   &\ZZ {XY}-velocity scaling factor ($>$ 0 for solid-body rotation) \\
\ZZ {VZROT}   &\ZZ {Z}-velocity scaling factor (not used if \ZZ {VXROT} = 0) \\
\ZZ {RBAR}    &Virial radius in pc (for scaling to physical units) \\
\ZZ {ZMBAR}   &Mean mass in solar units (for scaling) \\~\\
\ZZ {XCM}     &Displacement of subsystem (routine \ZZ {SUBSYS}; \# 17) \\
\ZZ {ECC}     &Eccentricity of relative motion for subsystem (\ZZ {ECC} $\le$ 1) \\
\hline
\end{tabular}
\end{table}
\medskip

   Four main types of input may be distinguished.
The integration parameters \ZZ {ETAI} = 0.03 and \ZZ {ETAR} = 0.06 usually
give reasonable results, although slightly smaller values are preferable.
Further discussions of time-step criteria and accuracy in the AC method
have been given elsewhere (Makino 1991a, Makino \& Aarseth 1992).
The optimum size of the neighbour sphere depends on the problem;
usually \ZZ {NNBMAX} $\simeq 10 + N^{1/2}$ is sufficient for small $N$.
For larger $N$ ($> 1000$), Makino \& Hut (1988) suggest
\ZZ {NNBMAX} $\simeq (N/8)^{3/4}$.
However, see Spurzem (1999) for a different point of view.

   An estimate of the initial neighbour radius (\ZZ {RS0}) for routine
\ZZ {NBLIST} should be consistent with the scaled density distribution.
For homogeneous systems \rn{contr} gives $R_s \, = \, (2 n_1 / N)^{1/3} \, R_h$,
where $n_1$ is the desired member\-ship (say \ZZ {NNBMAX}/2).
If necessary, the neighbour sphere is doubled or reduced by the volume factor
until the membership falls in the acceptable range [1, \ZZ {NNBMAX}].

   Another set of parameters specify the initial conditions.
The main decision-making is controlled by time intervals (\ZZ {DELTAT, ~TCRIT})
and error checking (\ZZ {QE}) for conservative systems.
Among the optional features are rotation, external potentials and the creation
of two initial subsystems.
To guard against typing errors, routine \ZZ {VERIFY}
checks some input parameters.

\section{NBODY2 Algorithms}
In the following subsections, we discuss some of the main procedures and
algorithms used by the code \ZZ {NBODY2}, including several optional features.

\subsection{Initial Conditions}
$N$-body simulations usually employ quite specific initial conditions.
However, for illustrative purposes
we include a brief discussion of routine \ZZ {DATA}.

   The present version provides a choice (option 5) between an artificial
system of constant density and a centrally concentrated Plummer model.
Local input parameters are first read (exponent $\alpha_s$,
maximum and minimum mass $m_1$ and $m_N$).
If $\alpha_s = 1$ or $m_1 = m_N$, all the individual masses are taken to be
unity, otherwise they are selected from a smooth power-law distribution,
where $\alpha_s = 2.35$ gives a Salpeter-type function.

   In the simplest case (\ZZ {KZ(5)} = 0), a uniform spherical system is
produced by choosing $x, \, y, \, z$ at random inside a unit sphere with
corresponding isotropic velocities having randomized magnitudes in [0,~ 1].
The second choice (\ZZ {KZ(5)} = 1) provides for a Plummer model with isotropic
velocity distribution
(cf.~Aarseth, H\'enon \& Wielen 1974 for a detailed algorithm).

\subsection{Scaling}
Results of $N$-body simulations are often presented in a variety of units
which hinders comparison with other work.
In the present code, a \hbox {convenient} set of units are introduced in routine
\ZZ {SCALE};
however, all scaling can be bypassed if desired (cf. option 16).
Once the initial conditions have been generated,
all coordinates and velocities are first expressed with respect to the centre
of mass rest frame.

   We introduce the so-called `standard' units recommended by Heggie \&
Mathieu (1986), in which $G = 1,\, M_0 = 1,\, E_0 = -1/4$, where $M_0$ is the
total mass and $E_0$ is the initial energy.
These units are suitable for most types of bound systems.
This choice corresponds to a virial equilibrium radius $R_v = 1$, rms velocity
$V_v = \sqrt 2 /2$ and
a mean crossing time
$T_{\rm cr} = 2 R_v/V_v = M^{5/2}_0 / (2 \, \vert E_0 \vert)^{3/2} = 2 \, \sqrt 2$.
The scaling to internal units is carried out in four separate stages:
\smallskip
\begin{enumerate}
\item Scale all the masses by the old total mass, $M^{\rm old}_0$,
to give $M_0 = 1$
\item Calculate the total kinetic, potential and virial energies
\item Scale the velocities by the specified virial theorem ratio (\ZZ {Q})
\item Rescale coordinates and velocities to yield the desired total energy.
\end{enumerate}

   Having transformed to scaled units, the output interval (\ZZ {DELTAT}) and
termination time (\ZZ {TCRIT}) are re-defined in terms of the crossing time.
Using the input parameters for the virial radius (\ZZ {RBAR}) in pc and mean
mass (\ZZ {ZMBAR}) in solar masses,
we also introduce scaling factors for conversion to physical units.
This gives rise to the time unit \ZZ {TSTAR} (in $10^6$ yr), velocity unit
\ZZ {VSTAR} (km s$^{-1}$) and a re-scaled mass unit \ZZ {ZMBAR} ($M_{\odot}$).

\subsection{Integration Algorithm}
We begin by summarizing the starting procedure of the AC method.
Given the initial conditions $m_i, \, {\bf r}_i, \, {\bf v}_i$ and a suitable
value of $R_s$, the two force polynomials are first initialized according to
\rn{fij} while the neighbour list is constructed.
We also form the total force \rn{ftot} and its first derivative, to be used for
coordinate predictions and neighbour corrections.
During the second stage, the next two orders \rn{f2ij} are calculated for each
component, using the neighbour list for identification.
The initialization procedure is completed by specifying irregular and
regular time-steps, $\Delta t_i$ and $\Delta T_i$, according to \rn{step},
whereupon the divided differences are formed by \rn{Ddiff}.
The main integration cycle consists of the following significant steps:

\begin{enumerate}
\item Select the next particle, $i$, to be advanced and define the
      current epoch by $\, t = t_i + \Delta t_i$
\item Make a new sorted time-step list
      if $t > t_L$ and adjust $\Delta t_L$
\item Compare $\, t + \Delta t_i \,$ with $\, T_0 + \Delta T_i \,$ to
      decide between a regular force
      prediction [case (1)] or a new total force summation [case (2)]
\item Predict coordinates of neighbours [case (1)] or all particles
      [case (2)]
\item Combine polynomials for $i$ and predict ${\bf r}_t$ and 
      ${\bf v}_t$ to order ${\bf F}^{(3)}$
\item Evaluate the irregular force ${\bf F}^{\rm old}_{\rm irr}$ and
      update the times $t_k$
\item Form new irregular differences and include the ${\bf D}^4$ term
\item $[$Case (1) only.$]$ Extrapolate the regular force and first derivative
      to give ${\bf F}_t$ and ${\bf F}^{(1)}_t$; proceed to step 15
\item Obtain the new irregular and regular force,
   ${\bf F}^{\rm new}_{\rm irr}$ and ${\bf F}^{\rm new}_{\rm reg}$, and form
      a temporary neighbour list. Check for $n_1 = 0$ or $n_1 > n_{max}$
\item Adjust the neighbour sphere $R_s$ and update the times $T_k$
\item Construct new regular differences and include the ${\bf D}^4$ term
\item Set ${\bf F}_t$ from \rn{ftot} and ${\bf F}^{(1)}_t$ using
\rn{taylor} for both types
\item Identify the loss or gain of neighbours and accumulate derivative
 corrections; update the neighbour list and convert to differences by \rn{Ddiff}
\item Prescribe the new regular time-step $\Delta T_i$
\item Update the new irregular time-step $\Delta t_i$
\item Repeat the cycle at step 1 until termination or output.
\end{enumerate}

  Here both force polynomials are integrated
to the same order; however, two separate time-step parameters are used
in \rn{step}.
Note that the decision to re-calculate the regular force is based on
the next {\it estimated} irregular time-step.
Force polynomials and their derivatives must be combined with care.
Thus the second regular difference required at step 5 (and at output times)
is obtained from a differentiation of \rn{taylor} at $t \not= T_0$ which yields
an extra term in ${\bf D}^3$.
Likewise for step 8, the regular force and its first derivative are 
evaluated to highest order at an intermediate time before the contributions
are added to the respective irregular parts.

   Various stratagems may be used to increase the efficiency of
determining the next particle to be advanced.
The earlier scheduling algorithm of searching a list of $\simeq N^{1/2}$
pre-selected members has been modified to include sequential sorting.
At time $t \geq t_L$ (with $t_L = 0$ \hbox {initially}),
a list is formed of all the
particles due to be advanced during the interval
$[t_L, \, t_L + \Delta t_L]$ and the list is then ordered sequentially.

   To include the case of repeated small time-steps and ensure that the global
time increases monotonically, we have adopted an insert procedure to maintain
the sequential ordering.
Thus the index of a particle satisfying
$t + \Delta t_i < t_L$, evaluated at the end of the cycle, is
inserted at the appropriate sequential location (routine \ZZ {INSERT}).
An updating of the sorted list is made if $t > t_L$, whereupon the interval
$\Delta t_L$ is stabilized on a membership chosen as a compromise
between the cost of sorting the quantities $t_j + \Delta t_j$ and inserting
a small number of particles in the sequential list.
This is achieved by employing two variable stabilization factors (in the
range $0.25 - 4.0$) which are used to determine the optimal membership
($\propto$ \ZZ {NNBMAX}).
In spite of this complication, comparison with the so-called `heap-sort'
algorithm advocated by Press (1986) favoured the present algorithm above
$N \simeq 100$, although the cost of both procedures is
$\propto N \,{\rm log}_2 N$.

\subsection{Force Polynomials}
At several stages, it is necessary to combine force polynomials and their
derivatives.
The prediction of the regular force is straightforward, using
\rn{Ft} with appropriate time arguments $t - T_k$ (cf. routine \ZZ {NBINT}).

   The first force derivatives are usually evaluated at the end of an
integration step of either type, since coordinates or velocities at a general
time $t \not= t_0$ or $t \not= T_0$ are obtained by an expansion with respect
to the interval $t - t_0$.
In the case of the irregular component, $t = t_0$ and {\rn{taylor} applies
directly; likewise for the regular component at an end-point when $t = T_0$.
However, the prediction of the regular force derivative at the end of a
sub-step which does not coincide with a total force calculation requires
special care.
Applying one time differentiation of \rn{Ft}, we obtain
\be
{\bf F}^{(1)}_t \, = \, {\bf D}^3 \, (t'_0 t'_1 + t'_0 t'_2 + t'_1 t'_2) \, + \,
{\bf D}^2 \, (t'_0 + t'_1) \, + \, {\bf D}^1,   \label{fdot}
\ee
where again $t'_k = t_0 - t_k$ denotes the relevant time intervals (since
$t_0 = t$ here).
Setting $t_k = T_k$ then yields the predicted regular force derivative which
is added to its irregular counterpart, obtained by \rn{taylor}, to form the
total force derivative, ${\bf F}^{(1)}$, employed in the subsequent predictions.

   The analogous expression for the second regular force derivative contains
an extra term $t'_0 = t_0 - T_0$ in the corresponding coefficient of
${\bf D}^3$ (cf.~\rn{taylor}).
It is required for the prediction of particle $i$ at the beginning of an
irregular step as well as at output and other high-order predictions.

   Force polynomials based on divided differences result in relatively simple
code.
However, the integration is most conveniently carried out in terms of
an equivalent Taylor series,
with (optimized) prediction to order ${\bf F}^{(3)}$ given by
\bea
{\bf r}_t \,&=&\, ((((\, 0.6 \,{\bf \widetilde F}^{(3)} \, t' \,+\,
{\bf \widetilde F}^{(2)}) \, {1 \over 12} t' \,+\, {\bf \widetilde F}^{(1)})
\, t' \,+\, {\bf \widetilde F}) \, t' \,+\,
{\bf v}_0) \, t' + {\bf r}_0 \nonumber \\
{\bf v}_t \,&=&\, (((\, 0.75 \, {\bf \widetilde F}^{(3)} \, t' \,+\,
{\bf \widetilde F}^{(2)}) \, {1 \over 9} t' \,+\, {\bf \widetilde F}^{(1)})
\, 1.5 \,t' \,+\, {\bf \widetilde F}) \, 2 \, t' \,+\, {\bf v}_0. \label{hipred}
\eea
Here the integration factorials in \rn{pred} and factorials in \rn{taylor}
are absorbed in ${\bf \widetilde F}^{(k)}$, and $t' = t - t_0$ represents the
integration interval.

   The coordinate and velocity increments of \rn{hipred} due to the corrector
${\bf D}^4$
contain four terms since all the lower derivatives are also modified in
\rn{taylor}.
Consequently, we combine the corresponding time-step factors for
${\bf r}_t$ and ${\bf v}_t$ to yield (in code notation)
\bea
s_6 \, &=& \, ((({\, 2 \over 3} \, t' \,+\, s_5) \, 0.6 \, t' \,+\,
s_4) \, {1 \over 12} \, t' \,+\, {1 \over 6} \, s_3) \, t' \nonumber \\
s_7 \, &=& \, ((\, 0.2 \, t' \,+\, 0.25 \, s_5) \, t' \,+\, {1 \over 3} \, s_4)
\, t' \,+\, 0.5 \, s_3,  \label{corr}
\eea
where again all factorials are absorbed in the force derivatives.
The coefficients are defined by $s_3 = t'_1 t'_2 t'_3, \,
s_4 = t'_1 t'_2 + t'_1 t'_3 + t'_2 t'_3, \, s_5 = t'_1 + t'_2 + t'_3$,
respectively, where the old definition of $t'_k$ still applies.
The extra factor ${t'}^2$ omitted from \rn{corr} is included
in ${\bf D}^4$.
The semi-iteration is applied to both polynomials here.
However, the regular corrections may not be beneficial if there are large force
derivatives, as in the case of small softening \hbox {parameters}
(i.e. $\epsilon < R_h/N)$.

   Although the AC scheme is based on the concept of nearest neighbours,
there are situations when this is less useful.
Thus the neighbour sphere determination
for a distant particle may lead to undesirable oscillations
and consequent time-step reduction.
This problem has been circumvented by including a nominal small central mass
in the irregular force if there are no neighbours inside a large
neighbour sphere ($R_s > 50 \, R_h$); hence adopting $n_1 = 0$
still permits large irregular time-steps.

\subsection{External Potentials}
We have included two optional external potentials, which are
treated in a similar way.
The first case (\ZZ {KZ(15)} = 1) is the Plummer potential
\be
\Phi_1 \, = \, - { G \, M_b \over (r^2 + a^2)^{1/2} },  \label{phi1}
\ee
where the background mass $M_b$ and length scale $a$
are specified as input (cf. \ZZ {XTPAR1, ~XTPAR2}).
The corresponding force is evaluated in routine \ZZ {XTRNL1} and added to the
regular component.
For initialization, the same routine adds the force and first derivative
to the regular terms.
The appropriate contributions are added to the potential and virial
energies (routine \ZZ {ENERGY}).

   We also include an external
logarithmic potential of the type
\be
\Phi_2 \, = \, { G \, M_g \over (R_g \, {\rm log}_2 (r / R_g) }, \label{phi2}
\ee
where $M_g$ is the mass scale and $R_g$ is the length scale
of the continuous mass distribution (\ZZ {KZ(15)} = 2).
A second routine (\ZZ {XTRNL2}) is constructed in a similar way as above,
with $C_g = M_g / R_g$ (denoted \ZZ {CGAS}).

\subsection{Inelastic Collisions}
An optional procedure (option 12) has been included for the coalescence
of two particles.
The neighbour scheme itself facilitates the determination of a collision
candidate since a search can be made during the irregular force loop.
Having identified a close neighbour,
the osculating semi-major axis, $a$, may be used in the collision criterion.
If accepted, the appropriate global index is saved until
the end of the integration cycle.
The main steps of the coalescence procedure (routine \ZZ {COAL}) are as follows:

\begin{enumerate}
\item Improve the collision candidate ($j$) to order ${\bf F}^{(3)}$
\item Obtain the potential energy with respect to both components
\item Form the two-body binding energy $E_b = -m_i m_j / 2 a$
\item Define new mass, coordinates and velocities using the index $\min\,(i,j)$
\item Calculate potential energy with respect to the composite body
\item Subtract the two-body energy and tidal potential energy from $E_t$
\item Reduce $N$ by 1 and update all \ZZ {COMMON} arrays
(routine \ZZ {REMOVE})
\item Predict coordinates and velocities of neighbours to order ${\bf F}^{(2)}$
\item Initialize force polynomials and time-steps for the new body.
\end{enumerate}

   This algorithm maintains fairly good energy conservation with respect to the
newly updated total energy ($E_t$).
Since the \ZZ {COMMON} arrays are addressed by global indices (rather than
pointers), the removal of variables associated with a given particle ($j$)
entails a compressing of arrays for all indices $k \geq j$, performed in
routine \ZZ {REMOVE}.
In addition, the neighbour lists require reduction by 1 for
all index references $> j$, as well as removal of the specified particle.

\subsection{Escaper Removal}
Routine \ZZ {ESCAPE} (option 13) contains procedures for the removal of escaping
particles.
A nominal tidal radius, $R_t$, is used as
distance criterion, with $R_t\,  = 10 \, R_h$.
The escape algorithm can be summarized as follows:

\begin{enumerate}
\item Select a candidate ($i$) for escape if 
$\, \vert {\bf r}_i \vert > 2 R_t$
\item Evaluate the binding energy
$E_i = {1 \over 2} {\bf v}^2_i + \phi_i ;\,$ accept escaper if $E_i > 0$
\item Check whether nearest particle is escaping or has large perturbation
\item Subtract $m_i E_i$ from the total energy $E_t$ and $m_i$ from $M_t$
\item Correct the irregular force and first difference for each neighbour
\item Reduce $N$ by 1 and update all \ZZ {COMMON} arrays
(routine \ZZ {REMOVE})
\item Re-define coordinates and velocities in the c.m. frame and correct $E_t$.
\end{enumerate}

   In a second stage, routine \ZZ {CMCORR} (option 18) adjusts all
coordinates and velocities to the centre-of-mass rest frame.
Appropriate modification of the total energy involves a correction to the
new kinetic energy, and the primary integration variables,
${\bf v}_0$ and ${\bf r}_0$ are updated consistently.

\subsection{Density Centre}
Models of centrally concentrated systems are most conveniently
analysed with respect to a well defined cluster centre, in accordance with
standard observational practice.
Following analytical estimates and extensive numerical experiments, Casertano
\& Hut (1985) have proposed an operational definition of the density centre
and corresponding core radius.

   The original definition of the core radius has been modified slightly in
order to obtain a convergent result using a smaller (central) sample
($n \simeq N/2$).
Thus the core radius is determined by the rms expression
\be
R_c = \left ( { \sum_{i=1}^n {\vert {\bf r}_i - {\bf r}_d \vert^2 \, \rho^2_i }
\over { \sum \rho_i^2} } \right )^{1/2},  \label{Rcore}
\ee
where ${\bf r}_d$ denotes the coordinates of the density centre and
$\rho_i = 15/(4 \pi \, r^3_6)$ is the density estimator defined with respect
to the sixth nearest particle, $r_6$.
We also generalize the
density estimator to include the actual mass of the {\it five} nearest
neighbours (Spurzem 1991) in order to be consistent with the original
expression for equal masses (cf. Eq. (V.3) of Casertano \& Hut).
The density centre and core radius are determined in routine \ZZ {CORE}
(option 8) according to the schematic algorithm:

\begin{enumerate}
\item Select all particles inside $\max \,(3 R_c, \, R_h)$ as data sample
\item Form a list of the six nearest neighbours for each selected particle
\item Sort the list of six square distances to find the largest ($r^2_6$)
\item Obtain the total mass ($M_5$) of the {\it five} nearest neighbours
\item Define individual mass densities
$\, \rho_i = 3 \, M_5/ (4 \pi \, r^3_6)$
\item Form the density centre
$\, {\bf r}_d = \sum {\bf r}_i \, \rho_i / \sum \rho_i$
\item Obtain core radius $R_c$ and average core density,
$<$$\rho$$> \,= \sum \rho^2_i / \sum \rho_i$.
\end{enumerate}

   In the AC scheme, central particles tend to have the largest neighbour
densities, $n_1 / R^3_s$; hence an appropriate averaging would also yield a
well defined density centre.
Where relevant, any discussions (and \hbox {algorithms}) \hbox {involving} the
central distance should be interpreted as referring to the \hbox {density} centre
(i.e. using $\vert {\bf r}_i - {\bf r}_d \vert$), which is updated at output times.

\subsection{Error Checking}
In standard simulations, energy conservation may be employed as a check
to safeguard the results from any spurious procedures such as force
discontinuities or inappropriate program usage.
Even systems which include collisions, escape and/or analytical external
potentials may be treated in this way, since the total energy is adjusted
accordingly.

   By saving all \ZZ {COMMON} variables at specified intervals (usually each
output), the last interval may be re-calculated with reduced time-steps if the
energy change exceeds the prescribed tolerance.
Since unit 1 (option 1) is reserved for \ZZ {COMMON} save at termination,
unit 2 is employed for this purpose.
Thus if \ZZ {KZ(2)} = 1, a \ZZ {COMMON} save is performed every output, provided
the relative energy error \ZZ {DE} (scaled by the kinetic or potential energy)
satisfies \ZZ {DE} $< 5\ast$\ZZ {QE},
where \ZZ {QE} is the specified tolerance.

   Modifications of the accuracy parameters take
place if \ZZ {KZ(2)} = 2.
Thus if \ZZ {DE} $> 5\ast$\ZZ {QE}, the previous \ZZ {COMMON} variables are
read from unit 2,
whereupon \ZZ {ETAI} and \ZZ {ETAR} are reduced by a factor of 2, and the
current time-steps are reduced partially (subject to $\Delta t_i \ge t - t_i$).
A second restart is carried out if this procedure does not satisfy the
accuracy criterion (using the counter \ZZ {NDUMP}).
If a restart is not required but \ZZ {DE} $>$ \ZZ {QE}, the integration
parameters are decreased by (\ZZ {QE/DE})$^{1/2}$,
whereas a small
increase is allowed if \ZZ {DE} $< 0.2\ast$\ZZ {QE} and \ZZ {ETAI} $<$
\ZZ {ETA0} (initial value of \ZZ {ETAI}).

\section{Practical Aspects}
The final section is devoted to practical aspects of $N$-body simulations,
and also summarizes some features of a test calculation.

\subsection{Performance and Accuracy}
The litmus test of a good code lies in its ability to produce satisfactory
solutions in a given time.
The CPU time requirement is particularly crucial for $N$-body simulations,
which are often designed to make full use of the available resources, as in
the case of dedicated workstations.
In view of the time-consuming nature of such calculations, code design must
inevitably be based on a trade-off between speed and accuracy.

   In the present version, we have made a compromise by defining the basic
variables ${\bf r}_0, \, {\bf r}_t, \, {\bf v}_0, \, t_0$ (and global time)
in full double
precision, with all other variables in standard precision.
This gives rise to a faster force calculation while the dominant terms of
the Taylor series are retained to higher accuracy at little extra cost.
Assuming a Fortran 77 compiler (or better), the precision of the code may
readily be changed by suitable declarations, taking care to match
the array sizes of the saved \ZZ {COMMON} blocks (routine \ZZ {MYDUMP}).
Moreover, on some types of hardware, the CPU time
may not depend significantly on the choice of precision.

   Actual CPU times are problem dependent and any empirical
relation is therefore only a rough guide.
In the absence of unique conversion factors for other machines, there is also
further uncertainty in estimating CPU times.
Timing tests have been carried out for Plummer models near equilibrium
over one crossing time with $N$ in the range [100, 2000] and softening
parameter $\epsilon = 4 \, R_v/N$.
A least square fit including all points gave a relation for the computing
time per crossing time as $T_c \propto \, N^{1.91}$ with a
correlation coefficient exceeding 0.9997.
Excluding $N = 100$ increased the exponent to 1.96.
This result may be compared with an older test for $N$ in [25, 250]
which gave a value of 1.6 (Aarseth 1985).

   The question of accuracy must also be considered, although it is not an
easy one.
It is well known that the numerical solutions for the point-mass
problem diverge on a relatively short time-scale (e.g. Miller 1964,
Goodman, Heggie \& Hut 1993).
However, in view of the chaotic nature of the $N$-body problem, it is not
justified to aim for the highest possible accuracy at the expense of
curtailing the integration time or particle number.
This is particularly relevant when using an
approximate model for the interaction potential.

   The most noticeable systematic errors are connected with the integration
of binaries, and this effect is also present when using a softening parameter.
We can measure the characteristic error of the two-body \hbox {motion} by studying
binaries in isolation, using the equivalent formulation with one force
polynomial (program \ZZ {NBODY1}).
For a binary with eccentricity 0.8 (and no softening) the systematic error in
semi-major axis per revolution is $\delta a / a = - 3 \times 10^{-5}$ with
$\eta = 0.02$ (or $- 6 \times 10^{-6}$ with $\eta = 0.01)$,
which corresponds to about 140 (200)  integration steps for each component.

\subsection{Softening}
The force law \rn{newton} is based on the softened potential
$- G \, m_j /(r_{ij}^2 + \epsilon^2)^{1/2}$ which was introduced
in order to model the interaction between galaxies (Aarseth 1963).
This potential represents the Plummer distribution,
where the softening size is related to the half-mass radius by
$R_h \simeq 1.3 \, \epsilon$.
We note that the maximum force occurs at $r = \epsilon / \sqrt 2$,
inside which the interaction tends to a harmonic oscillator.

   The introduction of a softened potential is a convenient artifact for
modelling larger systems by relatively small values of $N$, thereby
suppressing close encounter effects.
Considerations based on uniform systems in virial equilibrium lead to a
close encounter definition $r_{\rm cl} \simeq 4 \, R_h / N$ for a 90 degree
deflection between two typical particles.
On the other hand, the need for sufficient resolution places an upper
limit on the amount of softening for a given problem.
A general discussion of these aspects has been given by several authors
(e.g. White 1976, Gerhard 1981, Governato, Bhatia \& Chincarini 1991).
Although the present code has been designed for softened potentials, it
also works formally for the case $\epsilon = 0$ as long as there are no
critical encounters or persistent binaries of high eccentricity.

   In view of the slow asymptotic convergence of the standard softened
potential, it is desirable to consider expressions with a steeper
$r$-dependence in the transition region.
One such alternative is
$\phi_4 \,=\, - G \, m / (r^4 \, + \, \epsilon^4)^{1/4}$
which has been used for simulations of dwarf galaxies orbiting the Milky
Way (Oh, Lin \& Aarseth 1995).
Although the force evaluation is now more expensive,
the resulting force is a better
approximation to Newton's Law for $r > \epsilon$.
In the AC scheme, the steeper fall-off actually permits the point-mass
expression to be used for the regular force.
Implementation is straightforward, following identification of
all expressions involving $\epsilon^2$.

   The interpretation of the softening as an effective half-mass radius has
been subject to some confusion.
Although it has been traditional to generalize the interaction potential for
two unequal-mass bodies by replacing $\epsilon^2$ with
$\epsilon^2_i + \epsilon^2_j$ (e.g. White 1976), the corresponding force
does not describe correctly the motion of two overlapping Plummer spheres.
The actual force law of such interactions is rather complicated
(cf.~Wielen 1979), but alternatives have been discussed (Dyer \& Ip 1993).
However, this still leaves softened potentials as a useful tool for
suppressing two-body relaxation.
For example, collisionless simulations of collapsing systems may relate
softening to half-mass radius (cf. Aarseth, Lin \& Papaloizou 1988).

\subsection{Special Features}

   It is often desirable to perform large
simulations in several stages by saving all \ZZ {COMMON} variables on disc
or tape.
This can be done by specifying the run time at input (\ZZ {TCOMP}), together
with option 1.
Alternatively, on Unix machines the command `touch \ZZ {STOP}' creates a file
(\ZZ {STOP}) which triggers termination at the next timing check
(every \ZZ {NMAX} steps).

   A standard restart is made after reading the saved \ZZ {COMMON} file from
unit 1 (using \ZZ {KSTART} = 2).
For increased flexibility, some of the input parameters can be changed at
restart time.
Depending on the value of the control index, routine \ZZ {MODIFY} reads
one or two input lines (with the convention that only non-zero variables
are altered):

\begin{itemize}
\item {\ZZ {DELTAT, ~TNEXT, ~TCRIT, ~QE, ~J, ~KZ(J) ~~~~(KSTART} = 3 or 5)}
\item {\ZZ {ETAI, ~ETAR ~~~~(KSTART} = 4 or 5). }
\end{itemize}

   Among the various types of initial conditions, we include an example
of generating two separate subsystems (routine \ZZ {SUBSYS}, option 17) which
may be used for interacting binary experiments.
Here the second system is a replica of the first (after it has been scaled
in the usual way) and the two clumps are
assigned initial displacements ($\pm$\ZZ {XCM}) with specified orbital
eccentricity (\ZZ {ECC}), whereupon the particle number and total mass
are updated.
In this case, there is no well defined density centre, and the modification
of the neighbour sphere by a radial velocity factor outside the core should
be omitted (cf. option 10 in routine \ZZ {REGINT}).
Likewise, the determination of the density centre and the half-mass
radius should be modified accordingly (i.e. by using two separate procedures),
or suppressed altogether.

   The general algorithm of the AC scheme requires current values of all the
positions at each total force evaluation.
However, the frequency of full $N$ coordinate predictions may be reduced
considerably without unduly affecting the accuracy (option 14).
This is permissible because typical neighbours are predicted a number of times
at each irregular force summation, thus reducing the need for frequent
updating of all members.
Consequently, the full $N$ predictor loop is replaced by a neighbour
prediction if the membership $n_1$ {\it exceeds} the average neighbour number
(updated in \ZZ {KZ(14)} every output, provided the initial value is non-zero).
Another reduction of the numerical effort is made during initialization by
omitting distant contributions to the second and third force derivatives in
\rn{f2ij} (i.e.~$R > 3 \, R_s$).

\subsection{Data Analysis}
The emphasis of the code is on providing a versatile integration
tool, leaving the question of data analysis to the individual user.
The main facility here is the data bank (option 3) which
creates a file (unit 3) of the basic variables ($m, \, {\bf r}, \, {\bf v}$
and the identity) for each particle at output times \ZZ {TNEXT}
(frequency \ZZ {NFIX}).
This data may be read by a separate program, with the first record
(\ZZ {N, ~MODEL, ~NRUN}) containing the size (\ZZ {N}) of the
subsequent arrays.

   The general output on unit 6 (routine \ZZ {OUTPUT}) provides a summary of the
main diagnostics.
Most of this output refers to \ZZ {COMMON} variables.
Among the other quantities are the time expressed in crossing times (\ZZ {TC})
and in $10^6$ yr (\ZZ {T6}), half-mass radius ($<$\ZZ {R}$>$), maximum neighbour
density contrast (\ZZ {CMAX}), time-weighted neighbour number ($<$\ZZ {Cn}$>$),
centre-of-mass displacements (\ZZ {RCM, ~VCM}) and $z$ angular momentum (\ZZ {AZ}).

   Further optional diagnostic output is provided on unit 6 by routine
\ZZ {BODIES}.
Here an individual line is printed for each particle
(using option 9 to control the amount).
Likewise, a search for significant binaries is made at each output time or
at a specified frequency, \ZZ {NFIX} (\ZZ {KZ(6)} = 1 or 2).
Each output line contains the identity of components
and their masses, binding energy per unit mass, semi-major axis, mean motion,
separation, central distance, eccentricity and neighbour number.

Additional information about the radii of specified mass percentiles,
including the half-mass radius, is provided by routine \ZZ {LAGR} (option 7)
and is printed on unit 6 and/or 7, depending on the value of the option.
Further analysis or plotting routines may be called from
routine \ZZ {OUTPUT}, since all the current coordinates and velocities
(${\bf r, \, v}$) are known at this stage.

\subsection{Test Calculation}
It is instructive to illustrate the general workings of the code by
quoting a test calculation, although strict reproducibility may not
be achieved on different machines.
We choose input parameters for a collapsing system of
250 particles (cf.~test input file).
Some selected variables are displayed in Table~3 at every other output
time, using code notation.

\begin{table}
\caption{Summary of test calculation}
\begin{tabular}{rrrccccc}
\hline
   TIME&NSTEPI&NSTEPR&$<$NB$>$&Q&$<$R$>$&E&DE\cr
\hline
 0.0&0&0&  13&0.0& 1.87&--0.25008&0.000000 \cr
2.8&44387&5010&  13&0.38&1.27&--0.25008&  0.000007 \cr
5.7&227263&37293&7&0.59& 0.76&--0.25003& 0.000054 \cr
8.5&353397&60890&9&0.58&0.66&--0.25002& 0.000011 \cr
11.3&476792&83805&9&0.58& 0.65&--0.25001& 0.000005 \cr
14.1&601234&105861&10&0.62& 0.61&--0.25000& 0.000009 \cr
\hline
\end{tabular}
\end{table}
\medskip

   Comparison of the irregular and regular integration steps (Columns 2 and 3)
shows a ratio of about 6, for typical average neighbour numbers (Column 4)
of 10.
In spite of the violent relaxation, the total energy (Column 7) is quite
stable, with fairly small relative energy errors (Column 8).
The main error occurred at $t = 2 \, T_{\rm cr}$ when the time-step
parameters were reduced to 0.012 and 0.024 (from 0.02 and 0.04), respectively;
this was followed by a small increase.
Note that the half-mass radius (\ZZ {RSCALE}) should be updated fairly
frequently during phases of violent relaxation (cf.~\rn{contr}).

   The test calculation required a total CPU time of 35~s on a Sun UltraSparc~1
workstation, with clock speed 140~MHz and SPEC95 rating of 7.9.
Although the AC method gains in efficiency with respect to a single polynomial
formulation as $N$ increases, it can also be used for relatively small
systems.
Thus a second similar example with $N = 25$ conserved the total energy to
about $3 \times 10^{-5}$ (using \ZZ {NNBMAX} = 10 and $\epsilon = 0.25$)
after 10389 and 5134 irregular and regular steps, respectively.

\subsection{Program Modifications}
The variety of problems suitable for direct $N$-body simulations
is so large that one code cannot deal with all the requirements.
However, the basic solution method is sufficiently flexible for new
procedures to be included.

   Although the present code has been designed for scalar machines, it can
readily be modified to exploit special features available on
vector processors.
The main speed-up occurs for long loops, where the optimum
size is sometimes `quantized' to powers of 64 (e.g.~the CRAY supercomputer).
Thus it may be advantageous to increase the size of the neighbour array;
further discussion can be found elsewhere (Makino 1986,
Aarseth \& Inagaki 1986).
The total force loop itself should be simplified by omitting
the velocity-dependent part which becomes less effective with
increasing neighbour number.
Even for scalar machines, the implementation of a hierarchical time-step scheme
would be beneficial (McMillan 1986, Makino 1991b).

   Since the square root calculation is by far the most time-consuming part
of a direct $N$-body code, it may be desirable to consider alternative
procedures.
A fast algorithm based on a non-uniform look-up table for $1/r^3$ has proved
effective on scalar machines, where indirect addressing is not a problem
(cf. Aarseth, Palmer \& Lin 1993).
This algorithm would also be suitable
for the regular force calculation without significant loss of precision.

The code may readily be modified to include a population of massless test
particles.
Now all $N$ particles may be integrated in the standard way, with
the force loops performed over a smaller number of massive bodies.
When modelling close interactions, it is sometimes desirable to include
partially inelastic effects.
The collision determination would be similar to the coalescence
procedure but the velocity of each component should be modified by the
coefficient of restitution (Aarseth, Palmer \& Lin 1993).

   The effect of an external potential may also be studied as a perturbation
of the internal motions.
This description may be used for star clusters inside a galaxy or dwarf
galaxies orbiting a large galaxy.
In the case of open clusters, the assumption of circular motion near the
plane of \hbox {symmetry} \hbox {permits} simplified expressions for the
tidal force (cf. Aarseth 1985).
For more general types of orbits, it is convenient to employ local rotating
co\-ordinates with respect to a guiding centre (Oh, Lin \& Aarseth 1992).

   Cosmological modelling by direct N-body methods is limited to much smaller
particle numbers than can be studied by FFT or tree codes.
The standard AC code can be used for cosmological simulations as it stands
(i.e. \ZZ {Q} $>$ 1).
However, it is more efficient to employ comoving equations of motion.
Such a formulation has been described elsewhere (Aarseth 1985) and
is available as a separate program called \ZZ {COMOVE}.

   Finally, we mention the generation of
output for computer movies.
This requires monitoring an additional time-scale after each integration cycle.
All relevant particle coordinates can then be written to a separate file,
using scaled integer format of two bytes to save memory.
The actual movie production depends on available software; by now it is
possible to make nice movies on a PC using standard graphics packages.

\bibliographystyle{harvard}

\begin{thebibliography}{}

\bibitem[]{}
Aarseth, S.J., 1963, MNRAS, 126, 223.

\bibitem[]{}
Aarseth, S.J., 1985, in: Multiple Time Scales,
eds. J.U. Brackbill \& B.I. Cohen, (Academic Press, New York), p.~377.

\bibitem[]{}
Aarseth, S.J. \& Inagaki, S., 1986,
in: The Use of Supercomputers in Stellar Dynamics, eds. P. Hut \& S. McMillan,
(Springer-Verlag, New York), p.~203.

\bibitem[]{}
Aarseth, S.J., H\'enon, M. \& Wielen, R., 1974, Astron. Astrophys., 37, 183.

\bibitem[]{}
Aarseth, S.J., Lin, D.N.C. \& Papaloizou, J., 1988, ApJ, 324, 288.

\bibitem[]{}
Aarseth, S.J., Palmer, P.L. \& Lin, D.N.C., 1993, ApJ, 403, 351.

\bibitem[]{}
Ahmad, A. \& Cohen, L., 1973, J.Comput.Phys., 12, 389.

\bibitem[]{}
Casertano, S. \& Hut, P., 1985, ApJ, 298, 80.

\bibitem[]{}
Dyer, C.C. \& Ip, P.S.S., 1993, ApJ, 409, 60.

\bibitem[]{}
Gerhard, O.E., 1981, MNRAS, 197, 179.

\bibitem[]{}
Goodman, J., Heggie, D.C. \& Hut, P., 1993, ApJ, 415, 715.

\bibitem[]{}
Governato, F., Bhatia, R. \& Chinkarini, G., 1991, ApJ, 371, L15.

\bibitem[]{}
Heggie, D.C. \& Mathieu, R.D., 1986,
in: The Use of Supercomputers in Stellar Dynamics, eds. P. Hut \& S. McMillan,
(Springer-Verlag, New York), p.~233.

\bibitem[]{}
Makino, J., 1986,
in: The Use of Supercomputers in Stellar Dynamics, eds. P. Hut \& S. McMillan,
(Springer-Verlag, New York), p.~151.

\bibitem[]{}
Makino, J., 1991a, ApJ, 369, 200.

\bibitem[]{}
Makino, J., 1991b, PASJ, 43, 859.

\bibitem[]{}
Makino, J. \& Aarseth, S.J., 1992, PASJ, 44, 141.

\bibitem[]{}
Makino, J. \& Hut, P., 1988, ApJ Suppl., 68, 833.

\bibitem[]{}
McMillan, S.L.W., 1986,
in: The Use of Supercomputers in Stellar Dynamics, eds. P. Hut \& S. McMillan,
(Springer-Verlag, New York), p.~156.

\bibitem[]{}
Miller, R.H., 1964, ApJ, 140, 250.

\bibitem[]{}
Oh, K.S., Lin, D.N.C. \& Aarseth, S.J., 1992, ApJ, 386, 506.

\bibitem[]{}
Oh, K.S., Lin, D.N.C. \& Aarseth, S.J., 1995, ApJ, 442, 142.

\bibitem[]{}
Press, W.H., 1986,
in: The Use of Supercomputers in Stellar Dynamics, eds. P.~Hut \& S.~McMillan,
(Springer-Verlag, New York), p.~184.

\bibitem[]{}
Spurzem, R., 1991, Personal communication.

\bibitem[]{}
Spurzem, R., 1999, J. Comp. Applied Maths., 109, 407.

\bibitem[]{}
White, S.D.M., 1976, MNRAS, 177, 717.

\bibitem[]{}
Wielen, R., 1979, Mitt. Astron. Ges., 45, 16.

\end{thebibliography}

\end{document}